\documentclass[reprint,showpacs,showkeys,superscriptaddress,aps,pra]{revtex4-1}

\usepackage{graphicx}
\usepackage{epstopdf}
\usepackage{bm}
\usepackage{amssymb}
\usepackage{amsmath}
\usepackage{bbold}
\usepackage[colorlinks=true, linkcolor=blue]{hyperref}
\usepackage{ulem}
\usepackage{color, soul}
\usepackage{xcolor}
\usepackage{afterpage}
\usepackage{footmisc}
\usepackage{natbib}

\begin{document}

\title{High spin mixing conductance and spin interface transparency at $Co_2Fe_{0.4}Mn_{0.6}Si$ Heusler alloy and Pt interface}
\author{Braj Bhusan Singh}
\author{Koustuv Roy}
\author{Pushpendra Gupta}

\address{Laboratory for Nanomagnetism and Magnetic Materials (LNMM), School of Physical Sciences, National Institute of Science Education and Research (NISER), HBNI, Jatni-752050, India}
\author{Takeshi Seki}
\author{Koki Takanashi}
\address{Institute for Materials Research, Tohoku University, Sendai 980-8577, Japan}
\address{Center for Spintronics Research Network, Tohoku University, Sendai 980-8577, Japan}
 \author{Subhankar Bedanta}
 \email{sbedanta@niser.ac.in}
 \address{Laboratory for Nanomagnetism and Magnetic Materials (LNMM), School of Physical Sciences, National Institute of Science Education and Research (NISER), HBNI, Jatni-752050, India}
\begin{abstract}
Ferromagnetic materials exhibiting low magnetic damping ($\alpha$) and moderately high saturation magnetization are required from the viewpoints of generation, transmission and detection of spin wave. Since spin-to-charge conversion efficiency is another important parameter, high spin mixing conductance ($g_{r}^{\uparrow \downarrow}$) is the key for efficient spin-to-charge conversion. Full Heusler alloys e.g. $Co_2Fe_{0.4}Mn_{0.6}Si$ (CFMS), which are predicted to be 100$\%$ spin polarized, possess low $\alpha$. However, the $g_{r}^{\uparrow \downarrow}$ at the interface between CFMS and a paramagnet has not fully been understood. Here, we report the investigations of spin pumping and inverse spin Hall effect in $CFMS/Pt$ bilayers. Damping analysis indicates the presence of significant spin pumping at the interface of CFMS and Pt, which is also confirmed by the detection of inverse spin Hall voltage. We show that in CFMS/Pt the $g_{r}^{\uparrow \downarrow}$ (1.77$\times$10$^{20}$m$^{-2}$) and interface transparency (84$\%$) are higher compared to values reported for other ferromagnet/heavy metal systems.

\end{abstract}

\maketitle
\section{Introduction }
Spin transport across interfaces in ferromagnetic (FM)/heavy metal (HM) systems are important to develop future spintronics devices  [1,2]. Spin orbital torque  [3,4], spin transfer torque  [1,2], spin pumping/inverse spin Hall effect  [5$-$9], spin Seeback effects  [{10}$-${12}], etc. are major phenomenon which are predominantly affected by interface spin transport in FM/HM systems. Spin pumping is an efficient method to produce pure spin current ($J_s$), which is the flow of spin angular momentum, and investigate the spin propagation across FM/HM interfaces. The efficiency of spin transport at FM/HM interfaces is understandable by the factor known as effective spin mixing conductance ($g_{r}^{\uparrow \downarrow}$), which is related to $J_s$ by the expression[2].

\begin{equation}\label{equation1}
    J_s = \frac{\hbar}{4 \pi}g_{r}^{\uparrow \downarrow}\hat{m} \times \frac{d \hat{m}}{dt}
\end{equation}

where $\hat{m}$ is the unit vector of magnetization. $J_s$ can be converted into transverse voltage ($V_{ISHE }$) by inverse spin Hall effect (ISHE)  [13]:

\begin{equation}\label{eqiation2}
    V_{ISHE } \propto \theta_{SH} \Vec{J}_s \times \Vec{\sigma}
\end{equation}

where $\theta_{SH}$ is the spin Hall angle which defines the conversion efficiency between the charge current ($J_c$) and $J_s$, and $\Vec{\sigma}$ is the spin matrices governed by the spin polarization direction. Therefore, in order to get high $V_{ISHE }$ and hence $g_{r}^{\uparrow \downarrow}$  in a FM/HM heterostructure, $\theta_{SH}$ of the HM needs to be large. The value of $\theta_{SH}$ mostly depends on spin orbit interaction (SOI) and conductivity of the HM  [2,14]. Efficient interfacial spin transport critically depends on the type of interfaces and its associated FM and HM materials, FM materials with low magnetic damping ($\alpha$) are important to generate large spin current and hence high $g_{r}^{\uparrow \downarrow}$ [2]. In this context various low damping materials such as NiFe, CoFeB, and $Y_3Fe_5O_{12}$ have been studied. Further there is another class of half metallic materials e.g Heusler alloys which have been established as low damping systems. It is also noted here that spin pumping and the resultant spin current can be described as an accumulation of the up and down spins  [15]. Therefore, it is expected that in Heusler alloys spin pumping efficiency will be larger due to the presence of only one type of spins at the Fermi level. In addition, low magnetic damping and expected high spin pumping makes Heusler alloys suitable for the interface spin transport study and pure spin current-based spin torque nano-oscillators  [16]. There have been intense studies of spin dynamics with low damping materials, e.g. NiFe and CoFeB with various HM [17$-$21]. However, there are only a few reports on the spin dynamics of Heusler alloys with HM  [15,22]. $Co_2Fe_{0.4}Mn_{0.6}Si$ (CFMS) is a Heusler alloy which shows low damping and 100$\%$ spin polarization  [23]. Figure 1(a) shows a typical schematic for density of states for half metallic material. However, the spin pumping efficiency ($g_{r}^{\uparrow \downarrow}$) in CFMS/HM system has so far not been evaluated, which would help to understand its use for applications. Here, we report the spin pumping study in CFMS/Pt system with varying the thickness of Pt via (1) measurement of ISHE, (2) evaluation of effective mixing conductance $g_{r}^{\uparrow \downarrow}$  and (3) spin interface transparency.

\section{EXPERIMENTAL DETAILS}
The bilayer samples viz. S1$\#$ CFMS(20 nm)/Pt(3 nm), S2$\#$ CFMS(20 nm)/Pt(5 nm), S3$\#$ CFMS(20 nm)/Pt(7 nm), S4$\#$ CFMS(20 nm)/Pt(10 nm) and S5$\#$ CFMS(20 nm)/Pt(20 nm)  were prepared on MgO(100) substrates using dc magnetron sputtering in a vacuum system with base pressure $\sim$ 1 $\times$ $10^{-9}$ mbar  [24]. The prepared CFMS thin films were \textit{in-situ} annealed at $600^{\circ}$C/1hr to improve its crystallinity and surface quality. Reflection high energy electron diffraction (RHEED) patterns were acquired to characterize the surface and crystalline quality of CFMS layers. After the preparation of CFMS layer, Pt layer was deposited at room temperature by dc magnetron sputtering. The thickness of the films were evaluated using x-ray reflectivity (XRR) (data not shown). 
Ferromagnetic resonance (FMR) measurements have been performed in the frequency range 5-17 GHz on a coplanar wave guide in the flip-chip manner [25,26]. ISHE measurements have been performed by connecting a nanovoltmeter over two ends of the sample (sample size: 3 mm $\times$ 2 mm). The detail of the ISHE set-up can be found elsewhere [27].

\section{Results and Discussion}

\subsection*{Crystalline quality}
\begin{figure}[h!]
	\centering
	\includegraphics[width=0.45\textwidth]{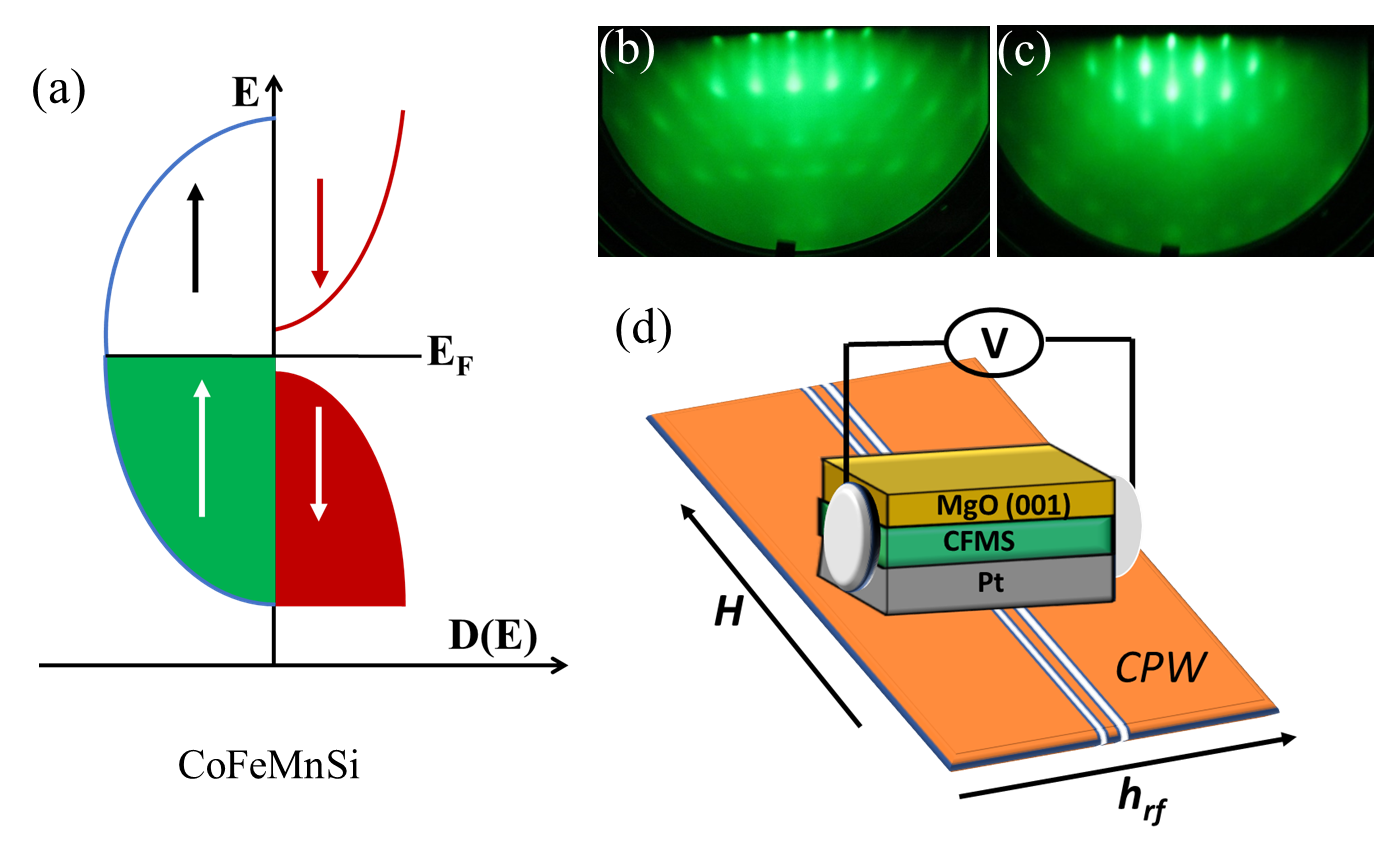}
	\caption{(a) A Schematic of density of states ($D(E)$) for an ideal half metal alloy. (b) and (c) are the RHEED patterns for the samples S4 and S5, respectively, on MgO (100) substrate in the [100] and [110] azimuths. (d) Schematic of the setup for ISHE measurement, where \textit{$h_{rf}$} is the \textit{rf} magnetic field generated in a coplanar wave guide (CPW) perpendicular to the applied magnetic field (\textit{H}).}
	\label{fig1}
\end{figure}

Fig. \ref{fig1} (b) and (c) show the RHEED patterns for the samples S4 and S5, respectively, observed in the MgO[100] and MgO[110] azimuths. From the streaks and spots of the RHEED patterns it is confirmed that the CFMS layer with the (001) crystalline orientation was epitaxially grown 
on the MgO (001) substrate. The streak lines which are elongated spots in vertical direction in the RHEED pattern implies the improvement of flatness at the CFMS surface. 

\subsection*{Magnetic Damping}
\begin{figure}[h!]
	\centering
	\includegraphics[width=0.45\textwidth]{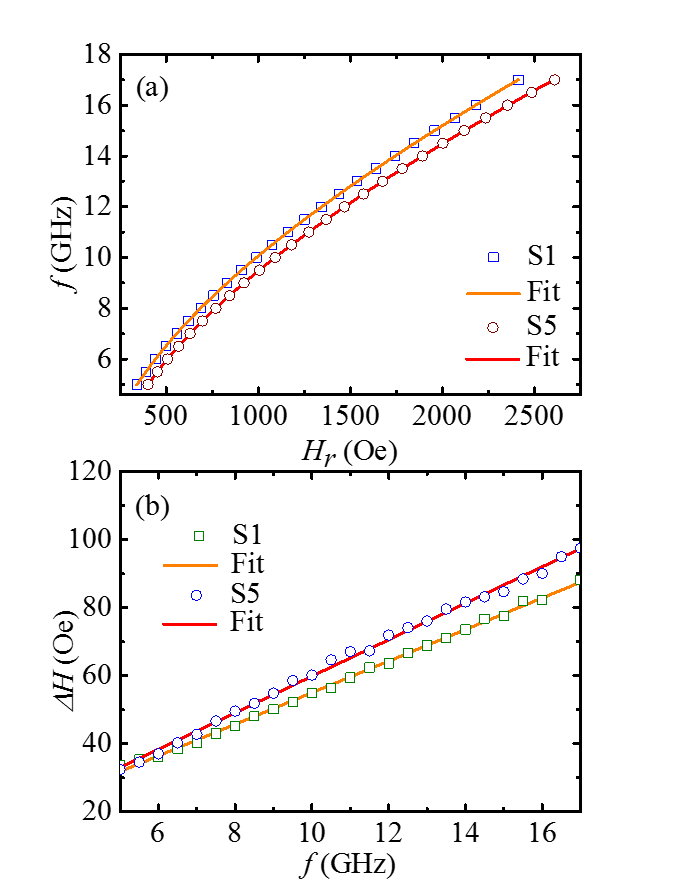}
	\caption{(a) $\textit{f}$ vs $H_r$ and (b) $\Delta H$ vs $\textit{f}$ plots for the samples S1 (open squares) and S5 (open circles). The solid lines are the best fits which are fitted by equations \ref{equation3} and \ref{equation5}. }
	\label{fig2}
\end{figure}

Fig. \ref{fig2}(a) and (b) show the plots of resonance frequency (\textit{f}) versus $H_r$ and $\Delta H$  versus \textit{f}, respectively. Here, the values of $H_r$ and \textit{$\Delta H$} were evaluated using FMR spectra (see fig. A1 in supplementary information). In order to evaluate the gyromagnetic ratio ($\gamma$) and effective demagnetization ($4\pi M_{eff}$), \ref{fig2}(a) was fitted to Kittel’s equation [28] given as:

\begin{equation}\label{equation3}
    \it{f}=\frac{\gamma}{2 \pi} \sqrt{(H_K+H_r)(H_K +H_r + 4 \pi M_{eff})}
\end{equation}

where 
\begin{equation}\label{equation4}
    4 \pi M_{eff} = 4 \pi M_s + \frac{2K_S}{M_st_{FM}}
\end{equation}

and $H_K$, $K_s$, $M_s$ $t_{FM}$, are anisotropy field, perpendicular surface magnetic anisotropy constant, saturation magnetization, and thickness of FM layer, respectively. $\alpha$ was evaluated by fitting data of Fig. \ref{fig2}(b) using the following expression [29]:

\begin{equation}\label{equation5}
    \Delta H = \Delta H_0 +\frac{4 \pi \alpha \it{f}}{\gamma}
\end{equation}

where \textbf{$\Delta H_0$} is the inhomogeneous broadening of linewidth which depends on the homogeneity of the sample. There are various other effects such as interface effect, impurity, magnetic proximity effects (MPE) etc., which also can enhance the value of $\alpha$ of the system. Hence, the total $\alpha$ can be written as: 
\begin{equation}\label{q6}
    \alpha = \alpha_{int} + \alpha_{impurity} + \alpha_{MPE} + \alpha_{sp}
\end{equation}

where $\alpha_{int}$ is the intrinsic damping, and $\alpha_{impurity}$, $\alpha_{MPE}$, and $\alpha_{sp}$ are the contribution from impurity, magnetic proximity effect (MPE), and spin pumping to the $\alpha$, respectively [30].

The linear behaviour of \textit{$\Delta H$} vs $\textit{f}$ plots implies the good homogeneity in our samples and it rules out the possibility of any kind of magnetic impurity of the FM layer.

\begin{table}[]
\begin{ruledtabular}
\caption{The values of $\alpha$ for 
        samples S1-S5}
\begin{tabular}{cc}

S1 & 0.0066 $\pm$ 0.0001 \\ \hline
S2 & 0.0063 $\pm$ 0.0001 \\ \hline
S3 & 0.0070 $\pm$ 0.0001 \\ \hline
S4 & 0.0085 $\pm$ 0.0001 \\ \hline
S5 & 0.0087 $\pm$ 0.0001 \\
\end{tabular}
\end{ruledtabular}
\end{table}

The values of $\alpha$ are larger than the reported value of single layer of CFMS ($\sim$ 0.004) [24]. This enhancement in the values of $\alpha$ is the indication of the spin pumping. However, we cannot rule out other effects e.g. MPE, and any impurities which may contribute in enhancing the value of $\alpha$ (see Eq. (\ref{q6})). In order to investigate the MPE or magnetic dead layer formation at the interface, we measured saturation magnetization (\textit{$M_s$}) for all the samples by SQUID magnetometer (data not shown). The measured values of \textit{$M_s$} for all the samples are found to be 861 emu/cc (S1), 842 emu/cc (S2), 792 emu/cc (S3), 845 emu/cc (S4) and 807 emu/cc (S5). The change in the values of Ms with the thickness of Pt may be due to MPE or dead layer formation  [31,32]. 

\subsection*{Inverse spin Hall effect measurement}
In order to confirm the spin pumping in our system, we performed ISHE measurements on all the samples as shown in schematic Fig. \ref{fig1}(d). The measurements are carried out at 11 mW power and 7 GHz frequency.

\begin{figure}[h!]
	\centering
	\includegraphics[width=0.45\textwidth]{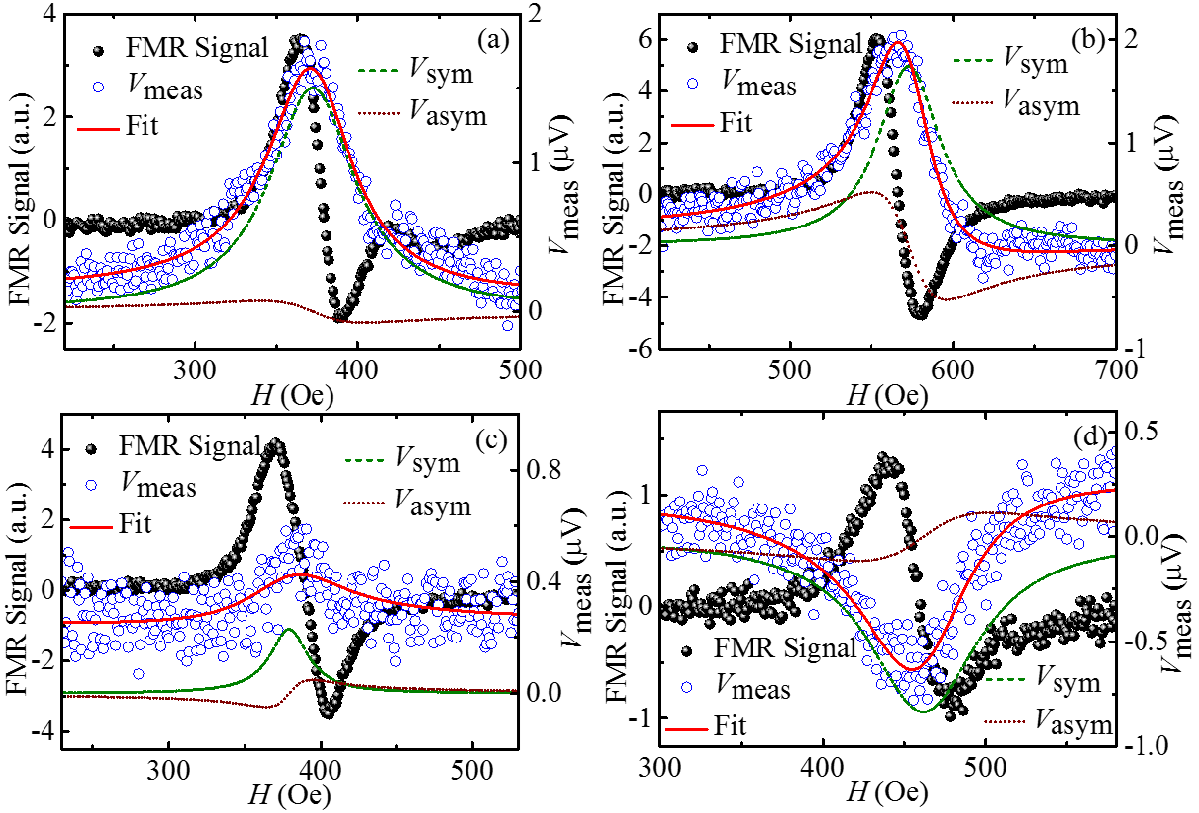}
	\caption{Voltage (\textit{$V_{meas}$}) measured across the sample with applied magnetic field along with FMR signal for sample S1 at the $\phi$ values of (a) 0$^{\circ}$, (b) 30$^{\circ}$, (c) 90$^{\circ}$, (d) 180$^{\circ}$. Open symbols are the experimental data. Solid lines are the fit to the experimental data using Eq. (\ref{q7}). Short dash and dotted lines are the symmetric ($V_{sym})$ and anti-symmetric ($V_{asym})$ components of the voltage}.
	\label{fig3}
\end{figure}

The angle $\phi$ denotes the angle between measured voltage direction and the perpendicular direction of applied DC magnetic field (\textit{H}). It has been previously found that CFMS thin films exhibit a cubic anisotropy  [33]. 

\begin{table*}[t]
\caption{Fitted parameters from $\phi$ dependent voltage measurements for all five samples }
\begin{ruledtabular}

\begin{tabular}{ccccc}

Sample & $V_{sp}$(V)$\times$$10^{-6}$ & $V_{AHE}$(V)$\times$$10^{-6}$ & $V_{AMR}^{\perp}$(V)$\times$$10^{-6}$ & $V_{AMR}^{||}$(V)$\times$$10^{-6}$ \\ \hline
S1     & 3.93 $\pm$ 0.05              & -1.14 $\pm$ 0.03              & 2.97 $\pm$ 0.06                       & 0.16 $\pm$ 0.03                    \\ \hline
S2     & 5.41 $\pm$ 0.16              & -1.28 $\pm$ 0.04              & 3.14 $\pm$ 0.18                       & 0.14 $\pm$ 0.04                    \\ \hline
S3     & 2.88 $\pm$ 0.06              & -0.78 $\pm$ 0.02              & 1.54 $\pm$ 0.07                       & 0.09 $\pm$ 0.02                    \\ \hline
S4     & 3.26 $\pm$ 0.05              & -0.67 $\pm$ 0.05              & 2.28 $\pm$ 0.06                       & 0.14 $\pm$ 0.03                    \\ \hline
S5     & 0.89 $\pm$ 0.02              & -0.64 $\pm$ 0.01              & 0.67 $\pm$ 0.02                       & 0.05 $\pm$ 0.01                    \\ 
\end{tabular}
\end{ruledtabular}
\end{table*}

Angle dependent measurements of the voltage have been investigated to remove spin rectification effects e.g. anisotropic magnetoresistance (AMR), anomalous Hall effect (AHE). Fig. \ref{fig3} shows the measured voltage (\textit{$V_{meas}$}) (open blue symbol) versus \textit{H} along with FMR signal (open black symbol) for sample S1 at the angles $\phi$= 0$^{\circ}$ (a), 30$^{\circ}$ (b), 90$^{\circ}$ (c) 180$^{\circ}$ (d). It should be noted that $\phi$= 0$^{\circ}$  means that the field was applied along the easy axis of the sample. There has been very less signal observed at $\phi$= 90$^{\circ}$ (Fig. \ref{fig3}(b)). This is due to the negligible amount of spin accumulation parallel to the applied magnetic field. It is evident from Fig. \ref{fig3}(a) and (d) that the sign of $V_{meas}$ is reversed when $\phi$ moves from  0$^{\circ}$ to 180$^{\circ}$. This indicates that the voltage is majorly produced by the spin pumping. It is well known that if the sign of the $V_{meas}$ does not reverse with angle, then the contribution solely comes from different spin rectification effects.

\begin{figure}[]
	\centering
	\includegraphics[width=0.45\textwidth]{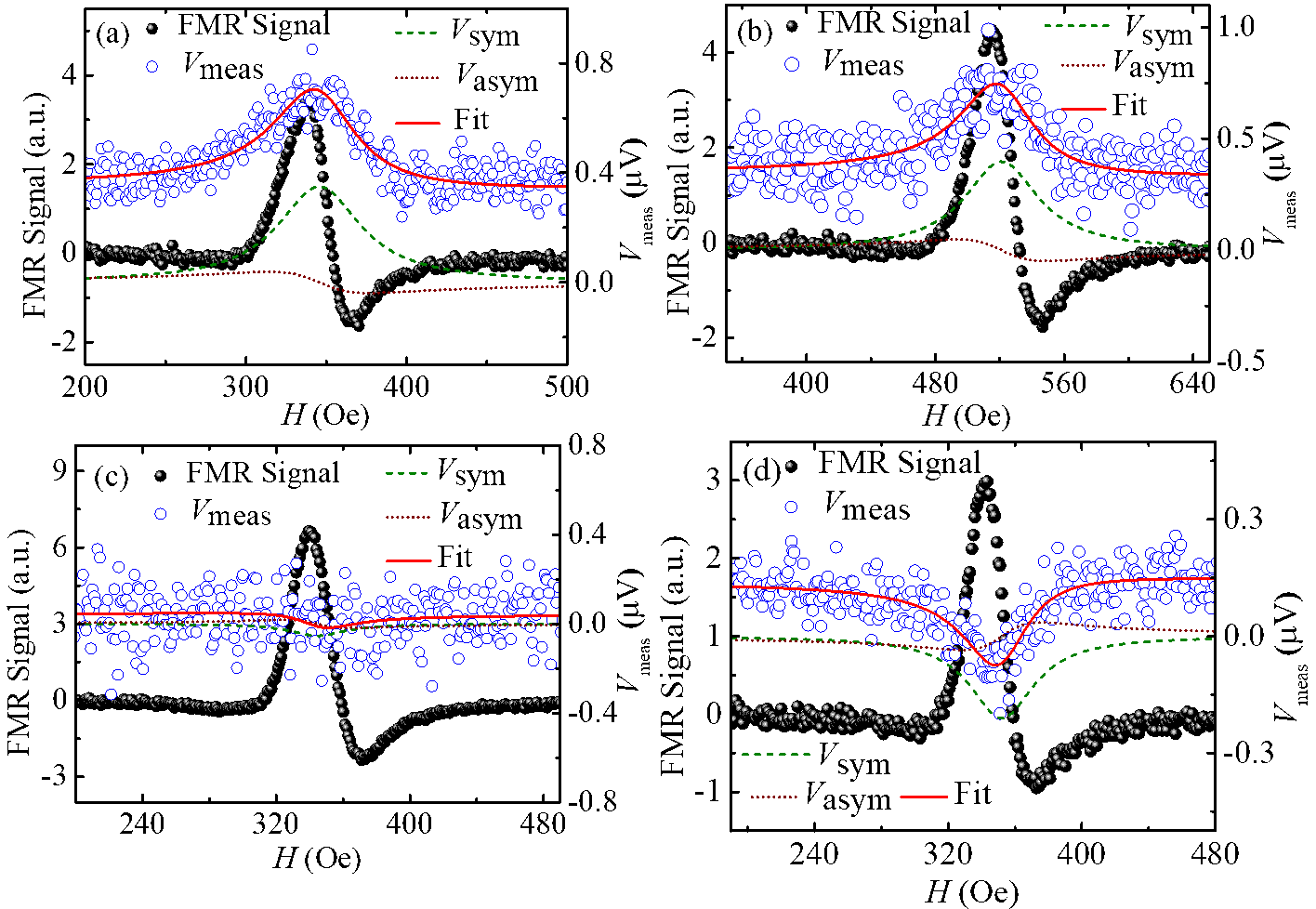}
	\caption{$V_{meas}$ versus $\it{H}$ and FMR signal for sample S5 at the $\phi$ values of (a) 0$^{\circ}$, (b) 30$^{\circ}$, (c) 90$^{\circ}$ (d) 180$^{\circ}$. Open symbols are representing the measured voltage. Experimental data are fitted (solid lines) using equation (\ref{q7}). Dashed and dotted lines are plots for components of symmetric ($V_{sym}$) and anti-symmetric ($V_{asym}$) voltage fitted to equation (\ref{q7}). }
	\label{fig4}
\end{figure}

Figure \ref{fig4} shows the \textit{$v_{meas}$} versus $\it{H}$ plot for the sample S5 measured at $\phi$= 0$^{\circ}$ (a), 30$^{\circ}$ (b), 90$^{\circ}$ (c) 180$^{\circ}$ (d). The similar kind of ISHE signal was observer for all the samples (data not shown). It is observed that the strength of the $V_{meas}$ for sample S5 (20 nm thick Pt) is three times smaller than that of the sample S1 (3 nm thick Pt).This is consistent to the fact that ISHE voltage is inversely proportional to the conductivity and thickness of the HM layer  [34]. 

For the separation of spin pumping contribution from the $\it{V_{meas}}$ by excluding other spurious effects, the $\it{V_{meas}}$ versus $\it{H}$ plots for the samples S1 (Fig. 3) and S5 (Fig. \ref{fig4}) were fitted with Lorentzian equation  [35] which is given by:

\begin{equation}\label{q7}
\begin{aligned}
    V_{meas} = V_{sym} \frac{(\Delta H)^2}{(H-H_r)^2+(\Delta H)^2}+ \\
    V_{asym} \frac{2 \Delta H (H - H_r)}{(H-H_r)^2+(\Delta H)^2}
    \end{aligned}
\end{equation}

where \textit{$V_{sym}$} and \textit{$V_{asym}$} are the symmetric and anti-symmetric components. Solid lines are fits to the experimental data. The $V_{sym}$   consists of major contribution from spin pumping, while minor contributions from AHE, and AMR effects. The AHE contribution is zero here if the \textit{rf} field and \textit{H} are perpendicular to each other, which is the case in our measurement. Whereas the AHE and AMR are the major contributions in the \textit{$V_{asym}$} component. Fig. 3 and 4 also show the plot of \textit{$V_{sym}$} (dashed line) and \textit{$V_{asym}$} (dotted line) separately for the samples S1, and S5, respectively.

\begin{figure}[htb]
	\centering
	\includegraphics[width=0.5\textwidth]{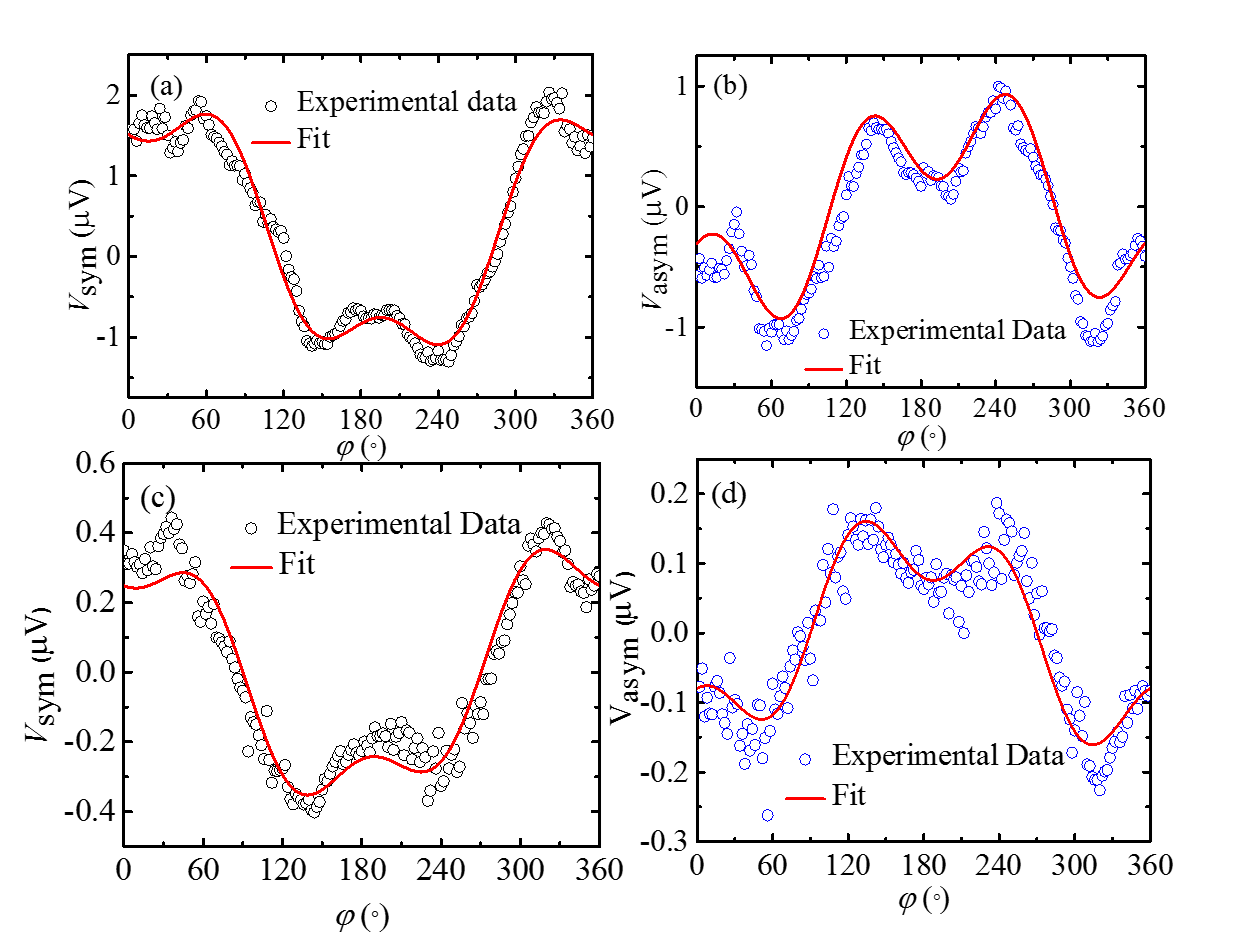}
	\caption{Angle dependent ($\phi$) $V_{sym}$ and $V_{asym}$ measurements for samples S1 (a and b) and S5 (c and d), respectively.}
	\label{fig5}
\end{figure}

In-plane angle dependent measurements of \textit{$V_{meas}$} were performed at the interval of 2\textit{$^{\circ}$} to quantify spin pumping and other spin rectification contributions (Fig. \ref{fig5}). It is a well-established method to decouple the individual components from the measured voltage  [30,36,37]. The model given by Harder \textit{et.al.}  [38] has considered the rectification effects i.e., parallel AMR (\textit{$V_{asym/sym}^{AMR ||}$}) and perpendicular AMR (\textit{$V_{asym/sym}^{AMR \perp}$}) to the applied \textit{rf} field and the AHE contribution due to the FM layer. The relation between the measured voltage and those rectification effects are as follows  [36]:

\begin{equation}\label{q8}
\begin{aligned}
 V_{asym}= V_{AHE}cos(\phi + \phi_0) sin (\it{\Phi})+ 
    \\V_{asym}^{AMR \perp} cos 2(\phi + \phi_0)sin(\it{\Phi})+ 
   \\ V_{asym}^{AMR ||}sin2(\phi + \phi_0)cos(\phi+\phi_0)
   \end{aligned}
 \end{equation}

\begin{equation}\label{q9}
\begin{aligned}
    V_{sym}= V_{sp}cos^3(\phi + \phi_0)+V_{AHE}cos(\phi + \phi_0)cos(\it{\Phi)}\\
    + V_{sym}^{AMR \perp} cos 2(\phi + \phi_0)cos(\phi+ \phi_0)\\
    + V_{sym}^{AMR ||}sin2(\phi + \phi_0)cos(\phi+\phi_0)
    \end{aligned}
\end{equation}

\textit{$V_{AHE}$} and \textit{$V_{sp}$} correspond to the AHE voltage and the spin pumping contributions, respectively. $\phi$ is the angle between applied \textit{H} and the \textit{rf} magnetic field which is always perpendicular in the measurement. The extra factor $\phi_{0}$ is taken to incorporate the misalignment of sample positioning in defining the $\phi$ value during the measurement. The detailed fits with and without incorporation of small offset in $\phi$ value is shown in Fig. A2 in the supplementary information. Further the AMR contribution also can be quantified by the following formula [36] :

\begin{equation}\label{q10}
  V_{AMR}=\sqrt{(V_{Asym}^{AMR \perp,||})^{2}+(V_{sym}^{AMR \perp,||})^{2}}  
\end{equation}

The \textit{$V_{Asym}^{AMR \perp,||}$} and \textit{$V_{sym}^{AMR \perp,||}$} are evaluated from the in-plane angle dependent \textit{$V_{meas}$} measurements by fitting those values by equations \ref{q8} and \ref{q9}, respectively. The extracted values of the various components are listed in the Table II.

 It is observed that the \textit{$V_{sp}$} is dominating over other unwanted spin rectification effects in all the samples. However, the magnitude of AHE is comparable to the spin pumping, which is decreased by one order for thicker Pt samples. It may be due to increase in conductivity of Pt layer due to increase in its thickness. It is well known that the AHE majorly depend on the magnetization of the sample due to berry curvature of the FM  [39]. The AHE contribution is an intrinsic property of the FM layer. The Co based FM materials are always a potential candidate for the AHE phenomena  [40,41]. The saturation magnetization measurements of all the samples, indicated the presence of MPE in Pt or dead layer formation at interface which may result in the decrease of \textit{$V_{AHE}$} contribution as the Pt thickness increases from 3 to 20 nm. However, the AMR values are of similar order in all the samples. The finite AMR contribution indicates that the samples are anisotropic in nature. The positive value of \textit{$V_{sp}$} indicates the positive spin Hall angle in Pt, which is consistent with literature [34]. 

\begin{figure}[]
	\centering
	\includegraphics[width=0.5\textwidth]{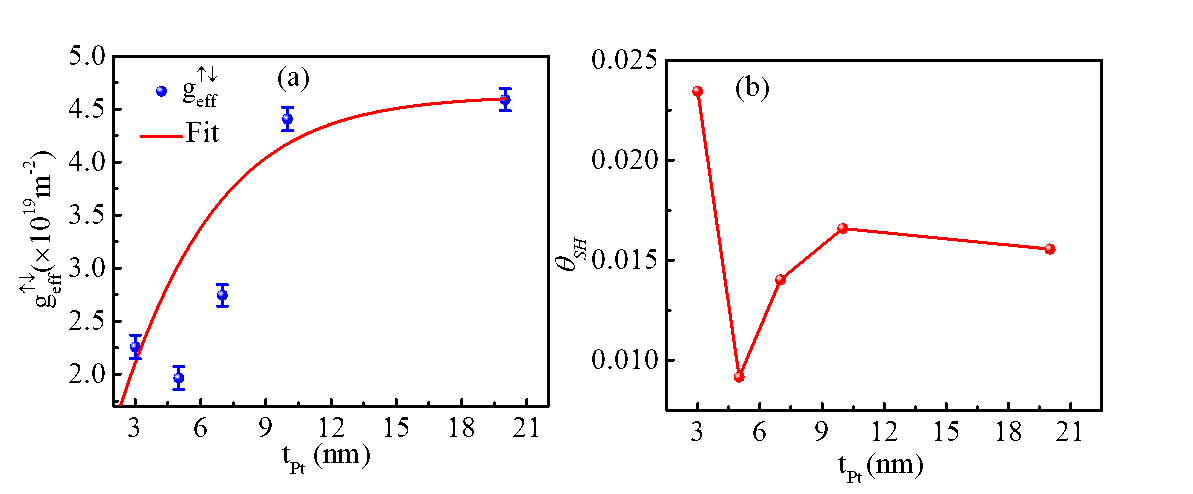}
	\caption{$g_{\it{eff}}^{\uparrow \downarrow}$ (a) and spin Hall angle (b) as a function of Pt thickness. Solid line in (a) is the best fit using the equation (12)}
	\label{fig6}
\end{figure}

The lowest $\alpha$ is found to be in S2 shows the maximum spin pumping voltage which is because of smooth interface between CFMS/ Pt. \textit{$V_{sp}$} is getting dominated by the conductivity of Pt for thicker Pt sample. Thus the \textit{$V_{sp}$} decreases with higher \textit{$t_{Pt}$} value.
Figure \ref{fig6} shows the graph between \textit{$g_{\it{eff}}^{\uparrow \downarrow}$} and Pt thickness. \textit{$g_{\it{eff}}^{\uparrow \downarrow}$}  was calculated by the following expression using damping constant[2]:

\begin{equation}\label{q11}
 g_{\it{eff}^{\uparrow \downarrow}}=\frac{\Delta\alpha 4\pi M_{s}t_{CFMS}}{g\mu_{B}} 
\end{equation}

where \textit{$\Delta\alpha$}, \textit{$t_{CFMS}$}, \textit{$\mu_{B}$}, \textit{g} are the change in the $\alpha$ due to spin pumping, the thickness of CFMS layer, Bohr magneton, Lande g- factor (2.1), respectively. In order to calculate the real part of spin mixing conductance \textit{$g_{\it{r}}^{\uparrow \downarrow}$}, we used the model which considered the spin memory loss (SML) mainly due to the interfacial roughness and disorder [44]. In this model effective spin mixing conductance is given by the following equation [44]:
\begin{widetext}
\begin{equation}\label{q12}
 g_{\it{eff}}^{\uparrow \downarrow}=
 \frac{r_{sl}cosh(\delta)+r_{sN}^{\infty}coth(\frac{t_{Pt}}{\lambda_{Pt}})sinh(\delta)}{r_{sl}[1+0.5\sqrt{\frac{3}{\epsilon}}coth(\frac{t_{Pt}}{\lambda_{Pt}}]cosh(\delta)+[r_{sN}^{\infty}coth(\frac{t_{Pt}}{\lambda_{Pt}})+0.5\frac{r_{sl}^{2}}{r_{sN}^{\infty}}\sqrt{\frac{3}{\epsilon}}]sinh(\delta)}
\end{equation}
\end{widetext}
where $\epsilon$ is the ratio of the spin conserved to spin flip relaxation times. According to [25], we set $\epsilon$ = 0.1 for the present Pt. \textit{$r_{sI}$}, \textit{$r_{sN}^{\infty}$}, \textit{$\delta$,$\lambda_{Pt}$}  are the interfacial spin resistance, the Pt spin resistance, the spin flip parameter for the CFMS/Pt interface, and the spin diffusion length in Pt, respectively. Fig. \ref{fig6} shows the fitting (solid line) of the data of \textit{$g_{eff}^{\uparrow \downarrow}$} using Eq. \ref{q12}. The fitting gives the values of \textit{$\lambda_{Pt}$} = 7.5 $\pm$ 0.5 nm, \textit{$g_{r}^{\uparrow \downarrow}$}= 1.77 $\pm$ 0.03 $\pm$ $10^{20}$ $m^{-2}$, and $\delta$ = 0.1. The values of \textit{$r_{sI}$} and \textit{$r_{sN}^{\infty}$} values are found to be 0.85 \textit{f} \textit{$\it{\Omega} m^2$} and 0.58 \textit{f} \textit{$\it{\Omega} m^2$}, respectively, which are  similar to the reported values for Co/Pt systems [42]. In CFMS/Pt system, the SML probability ([1-exp(-$\delta$)]$\times$100) is found to be 9.5 $\%$. It means the disorder at the interfaces is very small since spin depolarization is mainly caused by disorder at the interfaces. The interfacial spin resistance \textit{$r_{sI}$} is given by \textit{$r_b$}  / $\delta$ , where \textit{$r_b$} is the interface resistance. It indicates that most of spin current will flow through interface compared to bulk SOC of Pt if SML probability is large. However, in our case SML probability is very small (9.5$\%$), it means that most of the spin current is dissipating through bulk SOC of Pt, which produces charge current and hence to create \textit{$V_{ISHE}$}. Further, we compared the \textit{$g_{r}^{\uparrow \downarrow}$} and \textit{$g_{eff}^{\uparrow \downarrow}$} values evaluated in this work to the literature of various FM/HM systems in table III.

\begin{table}[]
\caption{The values of the spin diffusion length ($\lambda_{Pt}$), spin mixing conductance ($g_{r}^{\uparrow \downarrow}$), effective mixing conductance ($g_{eff}^{\uparrow \downarrow}$) from the literature and in this work.}
\begin{ruledtabular}

\begin{tabular}{cccc}
\begin{tabular}[c]{@{}c@{}}Layer\\   structure\end{tabular}                   & $\lambda_{Pt}$(nm) & $g_{r}^{\uparrow \downarrow}$($m^{-2}$) & $g_{eff}^{\uparrow \downarrow}$($m^{-2}$) \\ \hline
\begin{tabular}[c]{@{}c@{}}$NiFe/Pt$   {[}47{]}\end{tabular}                           & 7.7                         & 1.13$\times 10^{20}$                             & 3.02$\times 10^{19}$                               \\ \hline
\begin{tabular}[c]{@{}c@{}}$NiFe/Pt$    {[}6{]}\end{tabular}                            & 8.4                         & 2.53$\times 10^{19}$                             & 2.50$\times 10^{19}$                               \\ \hline
\begin{tabular}[c]{@{}c@{}}$CoFe/Pt$   {[}6{]}\end{tabular}                            & 7.6                         & 2.65$\times 10^{19}$                             & 2.50$\times 10^{19}$                               \\ \hline
\begin{tabular}[c]{@{}c@{}}$Co/Pt$   {[}6{]}\end{tabular}                            & 8.0                         & 1.42$\times 10^{19}$                             & 1.40$\times 10^{19}$                               \\ \hline
$Y_3Fe_5O_{12}/Pt$ {[}25{]}                                                             & 7.3                         & 3.90$\times 10^{18}$                             & 6.90$\times 10^{18}$                               \\ \hline
$CoFeB/Pt$ {[}26{]}                                                                     & 1.7                         & 3.90$\times 10^{19}$                             & 3.90$\times 10^{19}$                               \\ \hline
$Co_2MnSi/Pt$ {[}15{]}                                                                  & 10.0                        & ......                                           & 1.50$\times 10^{19}$                               \\ \hline
\begin{tabular}[c]{@{}c@{}}$Co_2Fe_{0.4}Mn_{0.6}Si/Pt$\\   {[}This work{]}\end{tabular} & 7.5                         & 1.77 $\times 10^{20}$                            & 4.21$\times 10^{19}$                               \\
\end{tabular}
\end{ruledtabular}
\end{table}

It can be observed from Table III that the values of \textit{$g_{r}^{\uparrow \downarrow}$} and \textit{$g_{eff}^{\uparrow \downarrow}$} are higher than the available literature values for the systems with Pt. Also, it should be noted here that the values of \textit{$g_{r}^{\uparrow \downarrow}$} and \textit{$g_{eff}^{\uparrow \downarrow}$} are large compared to the other reported low damping system viz. $Y_3Fe_5O_{12}/Pt$, $CoFeB/Pt$, and $Co_2MnSi/Pt$ [43,44]. Therefore, $CFMS/Pt$ system can be potential system for spin transfer torque and logic devices. In addition to the \textit{$g_{r}^{\uparrow \downarrow}$}, spin interface transparency (\textit{T}) is another parameter which is useful for spin-orbit torque-based devices. The value of \textit{T} is affected by the electronics structure matching of FM and HM layers. We used the following expression for the calculation of \textit{T}  [45]

\begin{equation}\label{q13}
    T = \frac{g_{r}^{\uparrow \downarrow} tanh(\frac{t_{Pt}}{2 \lambda_{Pt}})}{g_{r}^{\uparrow \downarrow} coth(\frac{t_{Pt}}{\lambda_{Pt}})+\frac{h\sigma_{Pt}}{2e^2\lambda_{Pt}}}
\end{equation}

where \textit{$\sigma_{Pt}$} is the conductivity of Pt layer. For \textit{$t_{Pt}$}=20 nm, \textit{T} is calculated to be 0.84 $\pm$ 0.02 by Eq.\ref{q13}, which is much higher than the values reported in the literature for $NiFe/Pt$ and $Co/Pt$ systems  [46]. Further it is also higher than the recent low damping $Co_2FeAl$/Ta layers system (68 $\%$)  [46]. It means that in $Co_2Fe_{0.4}Mn_{0.6}Si$/Pt system, matching of electronic structure is better than the other reported systems.
We also calculated the $\Theta_{SH}$ for the Pt using the following expression  [2]:

\begin{equation}\label{q14}
\begin{aligned}
    J_s \approx (\frac{g_{eff}^{\uparrow \downarrow}\hbar}{8\pi})(\frac{\mu_0 h_{rf}\gamma}{\alpha})^2\times\\
    [\frac{\mu_0 M_s\gamma+\sqrt{(\mu_0 M_s\gamma)^2+16(\pi f)^2}}{(\mu_0 M_s\gamma)^2+16(\pi f)^2}](\frac{2e}{\hbar})
    \end{aligned}
\end{equation}
\begin{equation}\label{q15}
    V_{ISHE}=(\frac{w_y}{\sigma_{FM} t_{FM}+\sigma_{Pt} t_{Pt}})\times \theta_{SH} l_{sd}^{Pt}tanh(\frac{t_{Pt}}{2 \lambda_{Pt}}) J_s
\end{equation}

The resistivity of the samples were measured using the four probe technique. The \textit{$\sigma_{Pt}$} and \textit{$\sigma_{CFMS}$} are found to be 2.3 $\times 10^{-7} \Omega$.m and 1.7$\times 10^{-6} \Omega$.m respectively. $\sigma$ corresponds to the conductivity of the individual layers. The \textit{rf} field ($\mu_0 h_{rf}$) and CPW transmission line width ($w_y$) value for our set up are 0.05 mT (at 11 mW rf power) and 200 $\mu$m, respectively. 
The obtained values of $\theta_{SH}$ are plotted in the Fig. \ref{fig6}(b). The values of $\theta_{SH}$ are comparable to the literature value [48]. In our case, we are observing higher SHA value for sample S1 in comparison to the sample S5. It may be due to the higher resistivity of the 3 nm Pt layers, which is consistent to the results obtained by J. Liu $\it{et al.}$  [47].

Further, we also performed power dependence of \textit{$V_{ISHE}$} measurements (Fig. A5, supplementary information). We observed a linear dependence of \textit{$V_{ISHE }$}, which confirmed the spin pumping at $CFMS/Pt$ interface.

\section{Conclusion}

We presented the study of the spin pumping and inverse spin Hall effect measurements in $CFMS/Pt$ bilayer samples. Angle dependent measurements of voltage were measured to quantify the various spin rectification effects. We observed a strong dependency of spin pumping voltage on the thickness of Pt. The spin pumping voltage was decreased when the thickness of Pt was increased, which may be due to increase in conductivity of Pt with thickness. The presence of substantial spin pumping keeps the damping constant values in the order of $\sim10^{-3}$. Spin mixing conductance ( \textit{$g_{r}^{\uparrow \downarrow}$} ) was obtained to be 1.77 $\times 10^{20}$ $m^{-2}$, which was higher than those for the other reported FM/Pt systems. In addition, we observed highest spin interface transparency (84 $\%$) compared to any other FM/Pt system. Low magnetic damping and large value of \textit{$g_{r}^{\uparrow \downarrow}$} with high interface transparency make the $CFMS/Pt$ system as a potential candidate for spintronic applications.

\section*{ACKNOWLEDGEMENT}

The authors acknowledge DAE and DST, Govt. of India, for the financial support for the experimental facilities. KR thanks CSIR for JRF fellowship. SB acknowledges ICC-IMR fellowship to visit IMR, Tohoku University, for this collaborative work to prepare the thin films. BBS acknowledges DST for INSPIRE faculty fellowship.

\section*{REFERENCES}
\begin{enumerate}
    \item S. D. Bader, and S. S. P. Parkin, \textit{Annual Review of Condensed Matter Physics} \textbf{1}, 71-88 (2010).
    \item Y. Tserkovnyak, A. Brataas, G. E. W. Bauer, and B. I. Halperin, \textit{Rev. Mod. Phys}. \textbf{77}, 1375-1421 (2005).

\item 	C. O. Avci, K. Garello, A. Ghosh, M. Gabureac, S. F. Alvarado, and  P. Gambardella,  \textit{Nature Physics} \textbf{11}, 570 (2015).
\item  Y.-T. Chen, S. Takahashi, H. Nakayama, M. Althammer, S.T.B. Goennenwein, E. Saitoh, and G. E. W. Bauer,  \textit{Phys. Rev. B} \textbf{87}, 144411 (2013).
\item	J. C. R. Sánchez, L. Vila, G. Desfonds, S. Gambarelli, J. P. Attan, J. M. De Teresa, C. Magn, and A. Fert,  \textit{Nature Communications} \textbf{4}, 2944 (2013).
\item	X. Tao, Q. Liu, B. Miao, R. Yu, Z. Feng, L. Sun, B. You, J. Du, K. Chen, S. Zhang, L. Zhang, Z. Yuan, D. Wu, and H. Ding,  \textit{Science Advances} \textbf{4}, eaat1670 (2018).
\item	B. Heinrich, C. Burroers, E. Montoya, B. Kardasz, E. Girt, Y.-Y. Song, Y. Sun, and M. Wu,   \textit{Phys. Rev. Lett.} \textit{107}, 066604 (2011).
\item	M. Obstbaum, M. Decker, A. K. Greitner, M. Haertinger, T. N. G. Meier, M. Kronseder, K. Chadova, S. Wimmer, D. Kdderitzsch, H. Ebert, and C. H. Back,   \textit{Phys. Rev. Lett.} \textbf{117}, 167204 (2016).
\item	K. Chen, and S. Zhang,   \textit{Phys. Rev. Lett.} \textbf{114}, 126602 (2015).
\item	K. Uchida, S. Takahashi, K. Harii, J. Ieda, E. Koshibae, K. Ando, S. Maekawa, and E. Saitoh,   \textit{Nature} \textbf{455}, 778–781 (2008).
\item	D. Meier, D. Reinhardt, M. van Straaten, C. Klewe, M. Althammer, M. Schreier, S. T. B. Goennenwein, A. Gupta, M. Schmid, C. H. Back, J.-M. Schmalhorst, T. Kuschel, and G. Reiss, \textit{Nature Communications} \textbf{6}, 8211 (2015).
\item	W. Lin, K. Chen,S. Zhang, and C. L. Chien,   \textit{Phys. Rev. Lett.} \textbf{116}, 186601 (2016).
\item	E. Saitoh, M. Ueda,H. Miyajima,  and G. Tatara,  \textit{ Appl. Phys. Lett.} \textbf{88}, 182509 (2006).
\item	Y. Tserkovnyak, A.  Brataas,  and G. E. W.Bauer,  \textit{Phys. Rev. Lett.} \textbf{88}, 117601 (2002).
\item	H. Chudo, K. Ando, K. Saito, S. Okayasu, R. Haruki, Y. Sakuraba, H. Yasuoka, K. Takanashi, and E. Saitoh, \textit{Journal of Applied Physics} \textbf{109}, 073915 (2011).
\item	V. E. Demidov, S. Urazhdin, G. de Loubens, O. Klein, V. Cros, A. Anane, and S.O. Demokritov,  \textit{Physics Reports} \textbf{673}, 1–31 (2017).
\item	S. Mizukami, Y. Ando,  and T. Miyazaki,   \textit{Phys. Rev. B} \textbf{66}, 104413 (2002).
\item	B. F. Miao, S. Y.  Huang, D. Qu,  and C. L. Chien,  \textit{Phys. Rev. Lett.} \textbf{111}, 066602 (2013).
\item	L. Chen,S. Ikeda, F. Matsukura,  and H. Ohno,  \textit{ Appl. Phys. Express} \textbf{7}, 013002 (2013).
\item  S.-I. Kim,  M.-S. Seo, Y. S. Choi,  and  S.-Y. Park,  \textit{Journal of Magnetism and Magnetic Materials} \textbf{421}, 189–193 (2017).
\item	M. Cecot, Karwacki, W. Skowroski, J. Kanak, J. Wrona, A. Ywczak, L. Yao, S. van Dijken, J. Barna, and T. Stobiecki,  \textit{Sci Rep} \textbf{7}, 1–11 (2017).
\item	I. Gościańska,  and J. Dubowik,  \textit{Acta Physica Polonica A} \textbf{118}, 851–853 (2010).
\item	A. Hirohata,  and K. Takanashi,  \textit{J. Phys. D: Appl. Phys.} \textbf{47}, 193001 (2014).
\item S. Pan, S. Mondal, T. Seki, K. Takanashi,  and A. Barman, \textit{Phys. Rev. B} \textbf{94}, 184417 (2016).
\item NanOsc AB. NanOsc AB Available at: http://www.nanosc.se/. (Accessed: 1st September 2019)	
\item B. B. Singh, S. K. Jena,  and S. Bedanta,  \textit{J. Phys. D: Appl. Phys.} \textbf{50}, 345001 (2017).	
\item	B. B. Singh, S. K. Jena, M. Samanta, K. Biswas, B. Satpati, and S. Bedanta, \textit{physica status solidi (RRL) – Rapid Research Letters} \textbf{13}, 1800492 (2019).
\item 	C. Kittel,  Phys. Rev. 73, 155–161 (1948).	
\item B. Heinrich, J. F. Cochran,  and R. Hasegawa,   \textit{Journal of Applied Physics} \textbf{ 57}, 3690–3692 (1985).	
\item A. Conca, S. Keller, L. Mihalceanu, T. Kehagias, G. P. Dimitrakopulos, B. Hillebrands, and E. Th. Papaioannou,\textit{ Phys. Rev. B } \textbf{93}, 134405 (2016).	
\item W. Amamou, I. V. Pinchul, A. H. Trout, R. E. A. Williams, N. Antolin, A. Goad, D. J. Ohara , A. S. Ahmed, W. Windl, D. W. McComb, and R. K. Kawakami,    \textit{Phys. Rev. Materials} \textbf{2}, 011401 (2018).	
\item X. Liang, G. Shi, L. Deng, F. Huang, J. Qin, T. Tang, C. Wang, B. Peng, C. Song, and L. Bi, \textit{Phys. Rev. Applied} \textbf{10}, 024051 (2018).
\item S. Mallick, S. Mondal, T. Seki, S. Sahoo, T. Forrest, F. Maccherozzi, Z. Wen, S. Barman, A. Barman, K. Takanashi, and S. Bedanta,   \textit{Phys. Rev. Applied} \textbf{12}, 014043 (2019).
\item K. Ando, S. Takanashi, J. Ieda, Y. Kajiwara, H. Nakayama, T. Yoshino, K. Harii, Y. Fujikawa, M. Matsuo, S. Maekawa, and E. Saitoh,  Journal of Applied Physics 109, 103913 (2011).	
\item 	R. Iguchi,  and E. Saitoh, \textit{J. Phys. Soc. Jpn.} \textbf{86}, 011003 (2016).	
\item 	A. Conca, B. Heinz, M.R. Schweizer, S. Keller, E. Th. Papaioannou, and B. Hillebrands, \textit{ Phys. Rev. B} \textbf{95}, 174426 (2017).
\item M. Harder, Y. Gui,  and  C.-M. Hu, \textit{Physics Reports} \textbf{661}, 1–59 (2016).	
\item  M. Harder, Z. X. Cao, Y. S. Gui, X. L. Fan,  and  C.-M. Hu,  \textit{Phys. Rev. B} \textbf{84}, 054423 (2011).	
\item 	N. Nagaosa, J. Sinova, S. Onoda, A. H. MacDonald, and N. P. Ong,  \textit{ Rev. Mod. Phys.} \textbf{82}, 1539–1592 (2010).	
\item W. Zhang, V. Vlaminck , J. E. Pearson. R. Divan , S, D. Bader, and A. Hoffmann,  \textit{Appl. Phys. Lett.} \textbf{ 103}, 242414 (2013).	
\item G. Zahnd, L. Vila, V. T. Pham, M. Cosset-Cheneau, W. Lim, A. Brenac, P. Laczkowski, A. Marty, and J. P. Attan,  \textit{ Phys. Rev. B} \textbf{98}, 174414 (2018).  
\item J. C. Rojas-Sánchez, N. Reyren, P. Laczkowski, W. Savero, J.-P. Attan, C. Deranlot, M. Jamet, J.-M. George, L. Vila, and H. Jaffrs,  \textit{Phys. Rev. Lett.} \textbf{112}, 106602 (2014).\\
\item 	H. Wang, Graduate dissertion, The Ohio state University (2015).	
\item 	M. Belmeguenai, K. Aitoukaci, F. Zighem, M. S. Gabor, T. Petrisor, R. B. Mos, and C. Tiusan,  \textit{Journal of Applied Physics} \textbf{123}, 113905 (2018).	
\item W. Zhang, W. Han, X. Jiang,  S.-H. Yang and  S. S. P. Parkin, \textit{Nature Physics} \textbf{11}, 496–502 (2015).
\item S. Akansel, A. Kumar, N. Behera, S. Husain, R. Brucas, S. Chaudhary, and P. Svedlindh, \textit{ Phys. Rev. B} \textbf{97}, 134421 (2018).	
\item 	J. Liu, T. Ohkubo, S. Mitani, K. Hono,  and M. Hayashi, \textit{ Appl. Phys. Lett.} \textbf{107}, 232408 (2015).	
\item H. Nakayama, K. Ando, K. Harii, T. Yoshino, R. Takahashi, Y. Kajiwara, K. Uchida, Y, Fujikawa, and E. Saitoh, \textit{Phys. Rev. B} \textbf{85}, 144408 (2012).

\end{enumerate}

\end{document}